\documentstyle[aps,prl,float,amsmath,amssymb,graphicx]{revtex}

\renewcommand{\vec}[1]{{\mathbf #1}}
\begin{document}
\draft
\preprint{UMIST/Phys/TP/00-10}

\title{Quantum Phase Transitions and the Extended Coupled Cluster Method}

\author{N.E. Ligterink\thanks{e-mail: ligterin@ect.it}\thanks{
Present address: ECT*, Strada delle Tabarelle 286, I-38050 Villazzano (Trento), Italy}%
, N.R. Walet\thanks{e-mail: N.R.Walet@umist.ac.uk}, and R.F. Bishop\thanks{e-mail: R.F.Bishop@umist.ac.uk}}
\address{Department of Physics, UMIST, PO Box 88, Manchester, M60 1QD,  UK}
\maketitle
\begin{abstract}
We discuss the application of an extended version of the coupled cluster
method to systems exhibiting a  quantum phase transition. 
We use the lattice O(4) non-linear sigma model in (1+1)-
and (3+1)-dimensions as an example.
We show how simple predictions get modified, leading to the absence of
a  phase
transition in (1+1) dimensions, 
and strong indications for a  phase transition in (3+1) dimensions.
\end{abstract}


The nature of the ground-state wave function for (infinite) quantum
systems is one of the underlying problems in many areas of physics,
from atomic and condensed matter to high-energy physics.  Special
problems are encountered when one studies the transition between
different ground states in cases where the balance of forces in a system with
competing interactions is shifted.  Since this is such a basic
problem, a large number of calculational schemes exist to study the
properties of the ground state. These schemes are either
imaginary-time approaches, based on the fact that in imaginary time
the state at lowest energy has the slowest rate of decay and can thus
be filtered out, or are real-time schemes, usually based on the
Rayleigh-Ritz variational principle. All of these methods have some
weaknesses. In high-energy physics, for example, imaginary-time
Monte-Carlo calculations can only be performed on finite lattices. As
a real phase transition can only occur in an infinite system, one has
to find ways to extrapolate to infinite lattices, and hope no
finite-lattice artifacts arise. Even if this can be done, an even more
fundamental problem in the Monte-Carlo integrations arises, due to the
uncontrollable fluctuations introduced by fermionic degrees of
freedom. This problem limits the validity of Monte-Carlo approaches in
condensed matter as well as those in lattice-(gauge)-field
theory. Real-time methods, on the other hand, are usually based on
rather complex many-body techniques, and lack the inherent simplicity
of the implementations of Monte-Carlo approaches. Also, even though
most real-time methods start from the variational principle, this
principle is often violated in an attempt to obtain a viable
calculational scheme.

An important difference between real- and imaginary-time approaches is
the way they deal with excitations -- whereas in imaginary-time
methods excited states need to be found by looking at exponentials
with faster decay than the ground state, or via complicated
Gramm-Schmidt orthogonalisation procedures, the calculation of
excitations is often much more straight-forward for the real-time
methods. Once we have constructed the ground state we can build
excited states on top of this ground state, using sets of excitation
operators. The calculational scheme used is then very similar for both
the ground state and the excited states.

In this letter we shall concentrate on zero-temperature, also called
quantum, phase transitions. The most important exact statement for
such situations is the Coleman no-go theorem \cite{Coleman} which
states that, in one space dimension and under rather generous
conditions, there is no phase transition to a state with long-range
order. As is well known from the exact solution of the $XXZ$ model in
(1+1)D \cite{Faddeev}, this does not preclude a phase transition with
an apparent gap-less mode, but without real long-range
order. Notwithstanding such special cases, the theorem is a strong
indictment against mean-field calculations, usually the lowest order
in the real-time approaches, since these often predict phase
transitions with long-range order even in one dimension.  The reason
is that mean-field theory ignores the strong interactions of the
massless modes that accompany such a phase transition, which pushes up
the mass of the would-be zero-mass modes.

One way to avoid this phenomenon is to concentrate on the phase with
unbroken symmetry, and to construct a calculational scheme for that phase.  Such
approaches have been used in spin-lattice systems and in lattice field
theories \cite{DamianRay,Krueger,Baker,US}, but it has been found that they
sometimes lead to termination points, probably indicating a phase
transition, for the (1+1)-dimensional system as well as in higher
dimensions. The excitation spectrum supports this behaviour, and
suggests a phase transition to a state with long-range order.

The most complicated part of the real-time approaches is the choice of
a truncation scheme. One must make a selection which has enough
degrees of freedom not to restrict the physical behaviour of the
system, nor to impose any assumptions of the nature of the phase
transition on the system, and not too many to make the calculation
infeasible. A good choice of wave function is key to any quantum
many-body calculation, and more importantly, there must be a way to
extend the calculations so as to test any assumptions being made.

Our approach is an example of such a formalism. The approach we use
is, however, systematic and will not give rise to spurious phase
transitions. The method is a recent offshoot of the coupled cluster
method, called the extended coupled cluster method (ECCM) \cite{ECCM}.
It has several advantages over earlier versions. First of all,
the mean-field approximation is a natural part of the hierarchy of
approximations inherent to the method.  It is thus suitable to deal
with states at both sides of a phase transition, where one exists. We
shall show in this letter that an (in principle infinite) set of
improvements removes the spurious transition in one dimensional
systems. For the moment we shall not include the mean-field term in
our calculations, but concentrate on two-site correlations instead.
While this restricts us to the symmetric phase, it makes it easier to
perform numerical calculations, which can be quite substantial.


We shall illustrate our method with a simple field theory,
the $O(4)$ non-linear sigma (NLS) model. We have chosen to discretise the
Lagrangian for this model on a lattice in space, keeping time
continuous. We then construct a Hamiltonian, which is the quantity
being studied. The basic degrees of freedom are unitary two-by-two
matrices, but these can also be represented by a four-dimensional unit
vector $(\vec{n},n_4)$, using a basis of Pauli matrices $\vec \tau$,
\begin{equation}
U = n_4 I + i \vec \tau \cdot \vec n.
\end{equation}
The Hamiltonian is
most succinctly expressed in the angle between nearest-neighbour
vectors, and the on-site generalised angular momenta,
\begin{equation}
H =
\frac{1}{2} \sum_{\vec{i}} \vec{I}_{\vec{i}}^2 + 
\lambda \sum_{\langle \vec{ i}\vec{j} \rangle}(
1-\cos \theta_{\vec{i}\vec{j}} ),
\label{fullhamil}
\end{equation}
where the sum over $\langle \vec{i}\vec{j} \rangle$ runs over all 
nearest-neighbour pairs, and counts each pair (or lattice link) only once.
The kinetic energy is proportional to the square of the 
(four-dimensional) angular momentum,  $\vec{I}_{\vec{i}}^2$.
This model shows great similarity with some
Heisenberg-like models used in condensed-matter physics.  

We expect a phase transition due to the competition between the
kinetic and potential terms. For small $\lambda$, the weak-coupling
limit, the first term is dominant and the system consists of a set of
independent rotors. In the strong coupling limit the potential
dominates and all vectors tend to align. As we increase $\lambda$ the
excitation spectrum is also expected to change from a free rotor
spectrum at small $\lambda$, to a spectrum containing three Goldstone modes,
corresponding to the dynamical breaking of chiral symmetry,
$
O(4) \rightarrow O(3),
$
at large $\lambda$.

This behaviour can most easily be seen in the mean-field
approximation.  Let us assume that on average all the vectors align
along the 4-axis. It can  be shown that in that case we obtain
\cite{OurNext} a non-linear version of the Mathieu equation, which has
a non-trivial solution (with broken symmetry) only for $\lambda D>3$,
with a standard first-order phase transition at $\lambda =3/D$, where
$D$ is the dimensionality of space. Below the phase transition point the
wave function at each site is just the spherically symmetric ``$S$-wave''
of the free rotor, and above the phase transition we find a two-dimensional
degeneracy, corresponding to positive and negative alignment with the
4-axis.

In the symmetric phase, one can ignore the one-body alignment, and
concentrate on two-site correlations. We have investigated these
processes using the normal coupled-cluster method \cite{US}, and have
found that without the one-body term there is a termination point for
one, two and three dimensions. This has been interpreted as a
signature of a phase transition. In one dimension this cannot be
true, since the Coleman theorem forbids such a phase transition. This
makes one distrust the other results as well. We show that the extended CCM 
is able to give much more reliable results.


In the extended coupled cluster method the bra- and ket-states are
parametrised independently, corresponding to the use of a bi-orthogonal
basis.  We define two operators $\hat S$ and $\hat S''$, from
which we  we build up a variational  functional $I$ which is
the ground state expectation value of the Hamiltonian in a special state,
\begin{equation}
I[S'',S] 
= \langle \Phi_0 | e^{\hat S''} e^{-\hat S} H e^{\hat S} | \Phi_0 \rangle=
\langle \tilde \Psi |  H  | \Psi \rangle\label{functio},
\end{equation}
where $|\Phi_0\rangle$ is the model, or reference, state, that must be
chosen separately.  A key ingredient in the definition of the
correlations is a set of generalised creation operators, that must be
chosen with reference to the state $|\Phi_0\rangle$, i.e., the
hermitean conjugates of the creation operators annihilate
$|\Phi_0\rangle$.  In the symmetric phase we
shall use the free-rotor vacuum, the product of ``$S$''-waves at each
lattice site, as reference state. It can be shown that the use of
single-site operators is a way to construct the mean-field
approximation.  When the mean-field is zero,
the lowest non-trivial correlations are two-site operators,
\begin{eqnarray}
\hat{S} & = & \sum_{\vec{i}, \vec{j}} \hat{S}_{\vec{i} \vec{j}} ,
\end{eqnarray}
where the sums run over the infinite lattice, and 
similarly for $\hat{S}''$.  The operators
$\hat{S}_{\vec{i} \vec{j}} $ 
create a simultaneous excitation at the lattice sites $\vec i$ and $\vec j$.
The precise definition of these operators is not crucial to the
current discussion, and can be found in our previous work
\cite{US}. By going to the coordinate representation, we find that
specification of the operators $\hat{S}_{\vec{i} \vec{j}} $
corresponds to specification of a single function
$S(\cos\theta_{ij})$.  We choose to work in this representation.  In the
numerical calculation we have to truncate up to a certain length of
the relative distance between ${\vec{i}}$ and ${\vec{j}}$. Both the
bra- and ket-state correlations have to be truncated in the same
manner to ensure the conservation of certain analytical properties of
the variational principle underlying the method.  The functions
$S(\cos\theta_{\vec{i}\vec{j}})$ can be expanded in Gegenbauer
polynomials of the relative angle $\theta_{\vec{i}\vec{j}}$ between
the unit vectors at lattice sites ${\vec{i}}$ and ${\vec{j}}$. Because
of the translational symmetry of the lattice the correlation functions
depend only on the relative vector ${\vec{i}}-{\vec{j}}$.  
For an optimal choice of $S$ and $S''$ the functional
$I$ must be stationary with respect to
independent variations of the bra- and ket-states. This leads to the
non-linear ECCM equations
\begin{equation}
\frac{\delta I[\{S'',S\}] }{\delta S''_{\vec i -\vec j}} = 0, \ \ \ \ \ \ 
\frac{\delta I[\{S'',S\}] }{\delta S_{\vec i - \vec j}} = 0.
\label{vareq}
\end{equation}
Since the ket state is defined as an exponential it contains
automatically all possible independent combinations of correlations on
top of the (lowest-order) sum of independent correlations. This is a
natural choice, since the correlation operators are combined in linked
objects, such as appear in Goldstone's linked cluster theorem
\cite{LinkedCluster}, and, indeed the coupled cluster method itself
provides by far the easiest proof of this important theorem
\cite{Bishop}.  Due to the similarity transform exponential $e^{-\hat
S}$ on the left-hand side of the Hamiltonian in the functional
Eq.~(\ref{functio}), $I$ is a polynomial in the correlation
functions. This has large advantages for high-order computer
implementations to solve the Eqs.~(\ref{vareq}).

We shall contrast the results of the extended coupled cluster method
to those of the normal coupled cluster method (NCCM), where the
bra-state is parametrised linearly in $S''$,
\begin{equation}
\langle \tilde \Psi | = \langle \Phi_0 | (1+ \hat S'') e^{-\hat S} .
\end{equation}
Therefore, in the NCCM, the overlap of the bra and ket states extends
only as far as the correlation operator $\hat S''$. Exponentiating
$\hat S''$ extends the overlap over the whole lattice and also breaks
certain symmetries of the functional which allows one to cross a phase
transition, instead of breaking down at the phase transition point,
which often occurs for the NCCM.


Using the time-dependent variational principle we can find low-lying
excitations of the system by considering harmonic fluctuations about
the ground state, also called the RPA approximation.  Since we have
the functional in an analytical form we are able to calculate the
excitation energies relatively straightforwardly.

\begin{figure}[htb]
\centerline{\includegraphics[height=9cm]{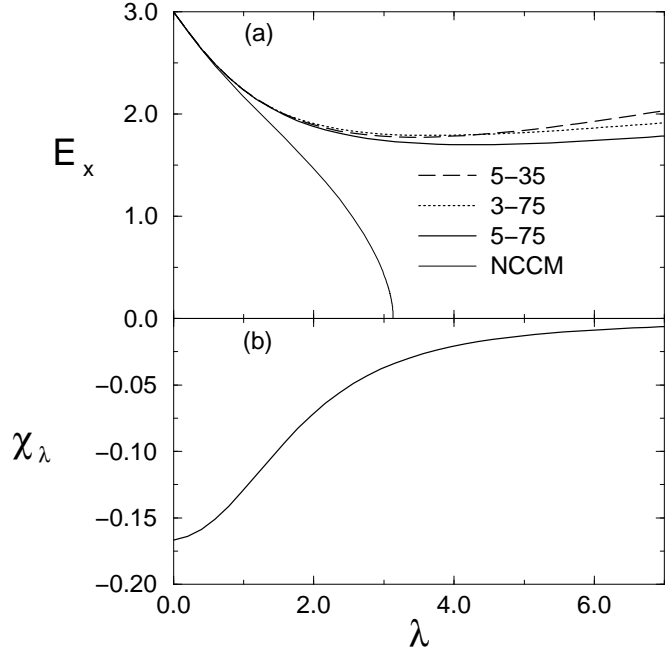}}
\caption{
(a) The lowest excitation energies and (b)  the second
derivative of the ground-state energy  for the NLS model in 1+1D. The labels $n$-$m$ refer to
truncations at $n$ base functions and $m$ correlations.}
\end{figure}
\begin{figure}[htb]
\centerline{\includegraphics[height=9cm]{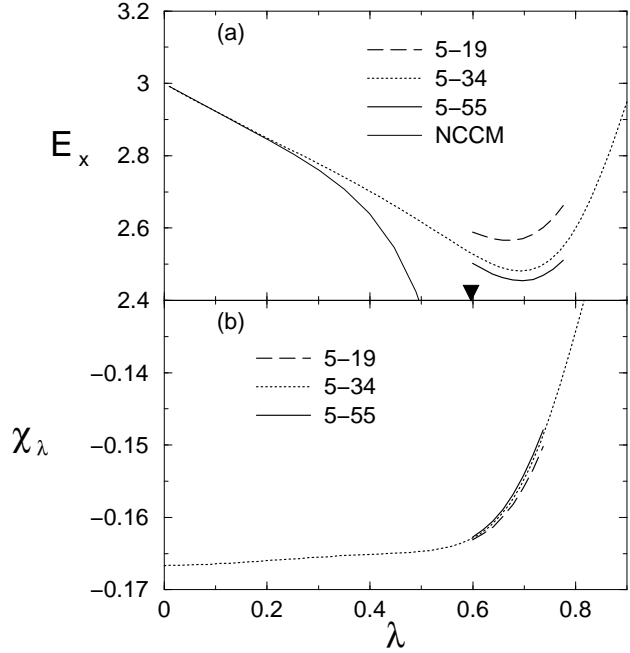}}
\caption{
(a) The lowest excitation energies, and (b)  the second
derivative of the ground-state energy  for the NLS model in 3+1D. The triangle indicates the
NCCM endpoint. The labels $n$-$m$ refer to truncations at $n$ base
functions and $m$ correlations.}
\end{figure}

We have applied the ECCM to the NLS model on a (1+1)- and a
(3+1)-dimensional cubic lattice.  We have studied various truncations,
and found that for the expansion of $S_{\vec i\vec j}$ in Gegenbauer
polynomials five such functions appear to be enough. For the
truncation of the range of the functions $S$ we have proceeded as far
as practicable, by including all correlations that fit within a cube
of size $L$. For the (1+1)-dimensional case we can almost use
arbitrarily large values of $L$, say including a hundred lattice sites. For
the $(3+1)$-dimensional result the best calculation was limited to those
correlations that fit inside  a cube of length 5.



For a linear chain the ground-state energy has converged completely
with correlations which extend over up to 75 lattice sites. The
susceptibility, the second-order derivative of the ground-state energy,
changes smoothly with  the coupling constant, indicating the absence of
a transition region. This is confirmed by the excitation energy. The
lowest excitation energy seems to have  converged  with 5 base
functions and 75 correlations, although the global trend sets in at much
lower orders. The mass gap remains finite. Clearly, these results are
in agreement with the Coleman theorem and are a step beyond
the mean-field, and NCCM, results.

For the cubic lattice the ground-state energy has more or less converged
for 5 base functions and all 55 correlations that fit in a $5\times 5
\times 5$ box.  The susceptibility shows two regions: a linear, almost
constant, susceptibility for the weak-coupling regime, and a steeper
behaviour for the strong-coupling regime. In low-order calculations
the behaviour between $\lambda \approx 0.6$ and $\lambda \approx 0.7$
changes drastically, while the trend on both sides outside this region
is already predicted by low-order results.  Even though the results do
not allow us the make a definite statement about the nature of the
transition, it seems weaker than a second-order phase
transition. The excitation energies again confirm these
results. Although the convergence is too slow to determine whether a
massless mode appears, it dips at the phase transition and rises fast
in the strong-coupling regime. Several low-lying excitation energies
turn complex beyond the phase transition, indicating that excitations
in this, distinctly different, phase are not properly described
without the one-site correlations necessary to describe symmetry
breaking. The absence of this phenomenon in (1+1)D is another
indication that no phase transition occurs for this case.

The slow convergence in the transition region also indicates that many
long-range correlations play an important role as is the case in a real
phase transition. The absence of this behaviour in (1+1)D is due to the
special properties of one-dimensional system that keep the correlation
length finite. The ECCM has the property that both the ket and bra states
are expressed in linked-cluster coefficients. Therefore, the length
over which ECCM correlation operators extend is a good indication of
the length of the physical correlations.

Since we have a dual expansion, in the extent of the correlations, and
in each correlation in a set of basis functions, we must look at the
truncation systematics for both of these.  For the (1+1)-dimensional
system, beyond a minimum length of the correlations (about 30 lattice
units), the results only depend on the number of basis functions,
which indicates a finite, physical correlation length.  The
convergence with the number of basis functions is quite fast,
indicating that the correlations are smooth functions of the relative
angle, which indicates that the large number of correlations are
essential to describe the phase transition properly since single 
correlation operators contribute only very little. 
%
%
This also holds for the (3+1)-dimensional system, but we never
fully reach convergence with the range of
the correlations. Both expansions seems to interplay in a complicated
way in the results. However, the ground-state energies generally
convergence much faster than the excitation energies.


With the ECCM we find signatures for a phase transition of the $O(4)$
non-linear sigma model in (3+1)-dimensions, which confirm our naive
expectations.  In (1+1)-dimensions the phase transition does not
occur, as expected from Coleman's theorem. These results show clearly
that ECCM is a proper many-body technique for the microscopic study of
phase transitions and other critical phenomena in cases inaccessible
with other methods, and we intend to use it in further studies.

We acknowledge support of a research grant (GR/L22331) from the
Engineering and Physical Sciences Research Council (EPSRC) of Great
Britain.

\end{document}